\documentclass[natbib]{svjour3}                     
\smartqed  
\usepackage{graphicx}
\journalname{Space Science Reviews}
\begin{document}

\title{Pressure and Ionization Balances in the Circum-Heliospheric
Interstellar Medium and the Local Bubble}

\titlerunning{Balances in the Local Bubble}
\author{Edward B. Jenkins}

\authorrunning{Jenkins} 

\institute{E. B. Jenkins \at
Princeton University Observatory,
Princeton, NJ, 08544-1001 USA \\
Tel.: 609-258-3826,
Fax: 609-258-1020,
\email{ebj@astro.princeton.edu}}

\date{Received: date / Accepted: date}

\maketitle
\begin{abstract}
A disconcerting mismatch of thermal pressures for two media in contact
with each other, (1) the warm, Circum-heliospheric Interstellar Medium
(CHISM) and (2) the very hot material within a much larger region called
the Local Bubble (LB), has troubled astronomers for over two decades.  A
possible resolution of this problem, at least in part, now seems
possible.  We now understand that earlier estimates for the average
electron density in the very hot LB plasma were inflated by an
unrecognized foreground contamination to the low energy diffuse X-ray
background measurements.  This foreground illumination arises from
photons emitted by charge exchange reactions between solar wind ions and
neutral atoms from the interstellar medium that enter into the
heliosphere.  However, with the resolution of this problem comes a new
one.  The high ionization fraction of helium in the CHISM, relative to
that of hydrogen, could be understood in terms of the effects from a
strong flux of EUV and X-ray radiation coming from both the Local Bubble
and a conductive interface around the CHISM.  A revision of this
interpretation may now be in order, now that the photoionization rate
from radiation emitted by hot gas the Local Bubble is lower than
previously assumed.

\end{abstract}
\begin{quote}
{\it ``When a thing ceases to be a subject of controversy,\\
it ceases to be a subject of interest''~} -- William Hazlitt (1778-1830)
\end{quote}

\section{Introduction: The Pressure Problem}\label{intro}

The history of science has highlighted many instances where the need to
resolve an incongruity in our perception of the natural world goaded us
into abandoning an entrenched idea -- a process that almost always has
led to an important new threshold for progress.  A major theme we will
address here is one such disparity, one which takes the form of an
apparent mismatch in thermal pressures\footnote{Throughout this paper,
pressures will be stated in terms of real pressures divided by the
Boltzmann constant, i.e., $p/k=nT$ for thermal pressures, because they
are easier to relate to particle densities and temperatures.} between
two phases of interstellar material that are in contact with each other. 
One such medium is the warm gas in the Circum-heliospheric Interstellar
Medium (CHISM)\footnote{Many interpretations in the recent literature
depict the heliosphere as being located close to the edge (but still
inside) a cloud of warm gas called the Local Interstellar Cloud (LIC)
that has coherent kinematics and is seen over much of the sky.  However
a recent investigation by Redfield \& Linsky  (2008) indicates that our
location is in a transition region between this cloud and another cloud
that is situated in the general direction of the Galactic center called
the G Cloud.  In recognition of this more complicated picture, we avoid
calling the material that surrounds us as the Local Interstellar Cloud,
since it now has a more restricted meaning, and replace it with the term
Circum-heliospheric Interstellar Medium (CHISM) that comprises both the
LIC and G Cloud.}, while the other is a much lower density medium that
surrounds the CHISM, which is understood to assume the form of a hot
plasma that resides within a large volume known as the Local Bubble
(LB).  There are several possible ways to confront this issue: (1)
acknowledge the presence of other forms of pressure (turbulent or
magnetic) whose differences can conceivably compensate for the thermal
pressure mismatch, (2) accept the possibility that a dynamical response
is underway and that this response will eventually resolve the
difference or (3) overcome the clash by revising our notions about the
physical properties of one or both regions.  We will touch upon each of
these three themes later in this article.

\section{Development of the Conventional View of the Local Bubble
(LB)}\label{LB}

The diffuse soft X-ray background detected by instruments on sounding
rocket flights in the 1970's was recognized by Williamson et al.  (1974)
to originate from hot plasmas in spaces within and beyond the Milky Way
[see also Burstein et al.  (1977)].  Soon afterward, it became clear
that much of this radiation at the lowest energies must come from an
irregularly shaped volume surrounding the Sun, since the X-ray energy
distribution showed little evidence for foreground absorption by cool,
foreground material, and in directions away from the Galactic plane the
intensities were anticorrelated with 21-cm emission by H~I, probably as
a result of the fact that the hot and cold gases displace one another 
(Sanders et al. 1977; Hayakawa et al. 1978; Marshall and Clark 1984).

From early evidence that a very limited amount of H~I was present out to
a radius of about 100~pc from the Sun [determined from various
investigations, summarized later by Paresce  (1984)], Fried et al. 
(1980) estimated that the thermal pressure of the X-ray emitting gas was
of order $p/k=10^4\,{\rm cm}^{-3}\,$K, a figure that was consistent with
more refined estimates that came later  (Marshall and Clark 1984; Cox
and Reynolds 1987; Snowden et al. 1990, 1998).  Determinations of the
average electron densities that were perhaps the most straightforward to
interpret came from measurements toward dense clouds whose distances
were known and which were expected to block the radiation.  A study of
such cases by Snowden, McCammon \& Verter  (1993), Kuntz, Snowden \&
Verter  (1997) and Burrows \& Guo  (1998) indicated that $n(e)\approx
0.005\,{\rm cm}^{-3}$; for an assumed temperature in the range
$1-2\times 10^6\,$K, this pointed toward a pressure $10^4 < p/k <
2\times 10^4\,{\rm cm}^{-3}$K.  Similar cloud shadowing measurements
conducted by Bowyer et al.  (1995) and Bergh\"{o}fer et al.  (1998), but
using EUV radiation instead of diffuse X-rays, indicated that
$p/k=19,500$ and $16,500\,{\rm cm}^{-3}\,$K, respectively.  While these
values for the thermal pressure inside the LB are higher than those
generally seen elsewhere in the ISM  (Jenkins and Shaya 1979; Jenkins,
Jura and Loewenstein 1983; Jenkins and Tripp 2001), they are not far
below an estimate of $2.8\times 10^4\,{\rm cm}^{-3}\,$K for the expected
{\it total\/} pressure in the Galactic plane  (Boulares and Cox 1990)
created by the weight of interstellar gases in the Galaxy's
gravitational potential.

A means for outlining the shape of the Local Bubble was developed by
Welsh et al.  (1999) and Sfeir et al.  (1999), who measured either the
sudden attenuation of EUV radiation or onset of strong Na~I absorption
features for stars at different distances and in different directions,
as sight lines started to penetrate the much denser gas beyond the
transparent hot medium.  As of now, the most refined picture of the
Local Bubble has been presented by Lallement et al.  (2003), who were
able to draw upon a larger number of stars than those available earlier.

\section{The Circum-heliospheric Interstellar Medium
(CHISM)}\label{CHISM}

Well before our awareness of the LB, early investigations of an
anisotropic backscatter of solar L$\alpha$ from outside the geocorona 
(Chambers et al. 1970; Bertaux and Blamont 1971; Thomas and Krassa 1971)
indicated that the solar system is moving through a low-density medium
containing some neutral hydrogen, consistent with a description made
shortly beforehand by Blum \& Fahr  (1969).  With more refined
observations and a theory of this backscattered radiation, coupled with
measurements of the absorption of EUV radiation from nearby stars 
(Vallerga 1996), there emerged a picture where the solar system is
moving through a partially ionized medium with a total particle density
$n({\rm H~+~He~+~e})\approx 0.35\,{\rm cm}^{-3}$ and  a temperature
$T=7000\,$K  (Bertaux et al. 1985; Frisch 1995, 2004; Slavin and Frisch
2002), which leads to a thermal pressure $p/k=2500\,{\rm cm}^{-3}\,$K. 
This material is similar in nature to that found in other gas complexes
in the local vicinity  (Lallement 1996; Redfield and Linsky 2004b,
2008).  As indicated in the introduction (\S\ref{intro}), the local
thermal pressure is in conflict with the much higher value for the
surrounding hot medium in the LB; this disparity has been a nagging
problem for many years and must be reconciled in some way.

\section{Attempted Solutions that Seem to Fail}\label{fail}

\paragraph{An imbalance is acceptable} At one time, when the age of the
LB and our location with respect to the outer boundary of the CHISM were
not very certain, one could imagine\footnote{As the author of this
article conjectured during a panel discussion at a conference on the
local interstellar medium, IAU Colloquium 81 in 1984.} that perhaps the
imbalance is real and, as a consequence, a shock front that started at
the perimeter of the CHISM is now approaching us from all directions. 
This shock should have a Mach number about equal to the square-root of
the relative pressure ratio, i.e., $v_{\rm shock}\approx 20\,{\rm
km~s}^{-1}$, and it has not yet reached us.  However, no evidence has
been found for such a front (i.e., from UV absorption lines, which are
very sensitive to small amounts of gas), and moreover we now know that
we are no more than a few pc away from the edge of the CHISM and that
the age of the LB is probably much longer than $10^5\,$yr.

\paragraph{Magnetic support} Since the local medium is partly ionized,
it should be strongly coupled to any magnetic field that could be
present.  For some time, the value and direction of the magnetic field
in the CHISM was unknown, so one could hold out some hope that magnetic
pressure could keep the CHISM from collapsing under the external
pressure from the LB.  Now, however, 3 independent studies using
different observational methods indicate that the magnetic field is
inclined relative to the upwind direction and creates an asymmetry in
the shape of the heliosphere, but it has a strength that ranges
somewhere between 1.8 and 2.5$\,\mu$G   (Lallement et al. 2005; Opher,
Stone and Gombosi 2007; Wood et al. 2007).  If the strength were as high
as 3$\,\mu$G, $B^2/(8k\pi)= 1700\,{\rm cm}^{-3}\,$K.  This much pressure
supplementing the thermal pressure still falls well short of the
apparent pressure of the LB.

\paragraph{Turbulent support}  If the LIC had a high level of
turbulence, it is conceivable that a random dynamical pressure $\rho
v_{\rm rms}^2$ could prevent it from collapsing under the external
pressure.  If we draw upon determinations based on other local clouds,
this seems unlikely.  Redfield \& Linsky  (2004b) found that on average
the thermal pressure was about 26 times larger than the pressure from
turbulence.

\section{Early Indications of Trouble with the X-ray Background
Measurements}\label{trouble}

By the time that the soft X-ray background started to be observed by
satellite missions, as opposed to the very brief sounding rocket
flights, it became apparent that enhancements in the background count
rates occurred during periods of increased solar activity  (Singh et al.
1987; Garmire et al. 1992; Snowden et al. 1994).  At the time, the real
cause of these increased event rates was not known; the presumption was
that they originated from the direct sensing of charged particles
associated with the solar events.  The investigators applied corrections
to the data to eliminate the time-variable components, but any rate
increases not identifiable with increased solar activity would have gone
unnoticed and would have contaminated the maps of X-ray background
intensity.

An important clue to the origin of the background arose from an
observation of EUV and X-ray emission ($0.1-2.0$~keV) from Comet
C/Hyakutake~1996~B2 reported by Lisse et al.  (1996), which was
interpreted by Cravens  (1997) to arise from the charge-transfer of
solar wind heavy ions as they interacted with the atmosphere of the
comet.  At a conference on the Local Bubble held in 1997, Freyberg 
(1998) and Cox  (1998) took note of the possibility that the solar wind
charge exchange (SWCX) mechanism could operate not only locally,
producing the X-ray enhancements clearly correlated in time with solar
activity, but also over the very large distances to the edge of the
heliosphere, where an interaction would occur with the incoming hydrogen
from the CHISM.  As a result, the signal would be so badly smeared out
in time that the contamination of the X-ray intensity maps would not be
recognized.  This contamination would have the consequence of misleading
us into thinking that the overall LB X-ray intensities, and thus the
emission measures\footnote{Emission measures (EM) denote the integral of
the square of the electron density along a line of sight for
temperatures that are relevant to the emission mechanism.} that
generally ranged from about 0.0018 to $0.0058\,{\rm cm}^{-6}\,$pc 
(Snowden 1998), are stronger than reality.  Later, Cravens  (2000)
stated, ``A simple model demonstrates that heliospheric X-ray emission
can account for about $25\% - 50\%$ of the observed soft X-ray
background intensities.''

While the need to understand the strength of the SWCX contamination is
crucial for a proper reassessment of the LB emission measures in various
directions, it is a complex problem to solve.  We have some
understanding of the types of processes that create energetic photons
when SWCX occurs  (Kharchenko and Dalgarno 2000; Pepino et al. 2004),
but there is nevertheless a multitude of factors that enters into any
attempt to correct the X-ray maps for this contamination.  One must
account for differences between the interactions with the fast and slow
solar winds (along with the occasional coronal mass ejections), the
differing geometrical distributions for different charge states as the
highly charged ions undergo successive charge exchanges, the changes in
the density of incoming H and He within the heliosphere, and so forth. 
Lallement  (2004) constructed a model for the emissivity of X-rays as a
function of location in the sky when the ROSAT\footnote{The mapping of
the soft X-ray background that came from the German {\it
R\"{o}ntgensatellit\/} (ROSAT) launched in 1990 is the most
comprehensive picture of the soft X-ray sky at the present time.} survey
was created, but initially with an uncertain global scaling factor for
the intensities.  She then derived this factor empirically by
determining the best fit of a polar plot of the corrected intensities to
the shape of the bubble's perimeter  (Lallement et al. 2003) and
concluded that a subtraction of the SWCX contribution could lower the
inferred thermal pressure by a factor 2.5, i.e., to as low as
$4000-6000\,{\rm cm}^{-3}\,$K.  A similar study by Bellm \& Vaillancourt 
(2005) arrived at a pressure reduction factor of 1.7.   These downward
revisions approached, but still did not equal, the pressure that seems
evident for the CHISM (\S\ref{CHISM}).

In the above paragraphs, we have followed a trail of reasoning that has
led to the conclusion that a simple interpretation of the apparent
intensities of the soft X-ray background is biased toward a value for
the emission measure in the Local Bubble that is too high.  This, in
turn, suggests that the average electron density (and hence the thermal
pressure) has been overestimated by a substantial factor.  One way to
overcome this systematic error is to measure the difference in
brightness between an opaque cloud at some known distance within the LB
and that toward a more distant screen in nearly the same direction, such
as the wall of the LB.  Here, an overall elevation of the flux by a
foreground emission source would have no effect.  This was the principle
behind the determinations of $n(e)$ using differential emission measures
of EUV radiation made by Bergh\"{o}fer et al.  (1998).   Unfortunately,
their investigation did not incorporate the knowledge (obtained later)
that the telescope they used on EUVE had some response to X-rays at
energies above 0.28~keV (B.~Welsh, private communication).  Since their
fiducial foreground blocking clouds (with $E_{b-y}=0.02$) had a
transmission of about 0.7 to such X-rays, it is possible that some of
the differential signal they sensed came from the X-rays, which could
have added to the decrement seen for the EUV radiation.

\section{How Else to Measure the Pressure in the LB?}\label{else}

\subsection{The Fine-Structure Excitation of Neutral Carbon}\label{C_I}

There are alternate means for sensing the pressure inside the LB,
although they are less direct than determining the emission measures in
either the EUV or soft X-ray bands.  One is to examine conditions inside
clouds other than the CHISM that are inside the LB.  If such clouds are
situated in front of bright, early-type stars and have a sufficient
internal density to show absorption features from C~I\footnote{The
dominant ionization stage of carbon is the singly ionized form.  The
fraction of these atoms that appear in neutral form, usually quite
small, is governed by the balance between photoionization and
recombinations with free electrons or negatively charged dust grains,
hence the absolute abundance of neutral carbon scales roughly in
proportion to the square of the local density.}, we can determine
allowed combinations of internal densities and temperature (or upper
limits thereof) by studying the relative strengths of absorption out of
different fine-structure levels of the ground electronic state.  The
populations of the upper levels are governed by an equilibrium between
collisional excitations (which scale with density) and spontaneous
radiative decays.

Jenkins  (2002) studied the C~I absorption features toward 4 stars that
were either inside or near the edge of the LB.  From a combination of
fine-structure population ratios and temperature constraints arising
from either line widths or thermal and ionization equibria, he concluded
that $10^3 < p/k < 10^4\,{\rm cm}^{-3}\,$K for the clouds under study.

\subsection{EUV Diffuse Background}\label{EUV_bkgnd}

Hurwitz, Sasseen \& Sirk  (2005) reported upper limits obtained from
observations by the {\it CHIPS\/} spectrometer\footnote{The {\it Cosmic
Hot Interstellar Plasma Spectrometer\/} ({\it CHIPS}) was launched in
early 2003 to measure the strengths of emission lines from the LB hot
plasma over the wavelength range $90-265\,$\AA.} for the emission of
radiation from highly ionized iron atoms in collisional ionization
equilibrium within the LB.  If the abundance of iron conformed to the
solar abundance ratio to hydrogen, their upper limit for this emission
at moderate Galactic latitudes indicated emission measures ${\rm
EM}<0.0004\,{\rm cm}^{-6}\,$pc (at a 95\% confidence level) for plasmas
in the temperature range $10^{5.55}<T<10^{6.4}\,$K.  This limit is about
an order of magnitude below the typical EM values seen in the soft X-ray
background without corrections for the SWCX emission, and it is
consistent with an upper limit for EUV iron-line emission detected by
{\it ALEXIS}\footnote{The {\it Array of Low Energy X-ray Imaging
Sensors} ({\it ALEXIS}) satellite experiment was launched in 1993 and
observed radiation in narrow energy bands centered on 186, 172, and
130\AA.} (Bloch et al. 2002).  Evidently, the relative SWCX
contamination for the Fe emission lines is smaller than for the soft
X-ray emission.

A troubling issue with the Fe-line measurements is the possibility that
the iron atoms that should accompany the gas may be depleted onto dust
grains, thus causing the abundance of these ions relative to protons in
the plasma to be below the solar abundance ratio.  While one might
conclude that the constraint on the EM determined by {\it CHIPS\/} could
be weakened considerably by this depletion, it appears unlikely that
this is happening: observations of X-ray absorption features from highly
ionized Fe and Ne elsewhere within the Galaxy by Yao et al.  (2006)
indicate a normal abundance ratio in the hot ISM.

\subsection{Cloud Lifetimes}\label{lifetimes}

Warm clouds immersed in a hot medium should have their outer layers
heated by conduction.  If radiative losses are small, this heating will
cause an evaporation of atoms on the cloud's surface into the
surrounding hot medium.  (If the radiative losses are large, then the
reverse process of condensation will take place instead.)  In the
absence of a magnetic field that could inhibit conduction by some modest
factor, the characteristic time scale for the shrinkage of a cloud due
to a steady erosion by evaporation is given by
\begin{equation}\label{evap_time}
M/\dot{M}=0.65np_4^{-5/6}r_{\rm pc}^{7/6}\,{\rm Myr}
\end{equation}
 (Slavin 2007), where $n$ is the cloud's internal gas particle density,
$p_4$ is the thermal pressure in units of $10^4\,{\rm cm}^{-3}\,$K, and
$r_{\rm pc}$ is the cloud's radius in pc.  If $p_4=0.2$, a cloud with
$n=0.2\,{\rm cm}^{-3}$ and $r_{\rm pc}=2$ (i.e., the CHISM or similar
clouds in the local vicinity) should have a shrinkage lifetime of about
1~Myr, but if $p_4=1.5$, this characteristic time decreases to only
0.2~Myr, which becomes inordinately short compared to the estimated age
of $10-15\,$Myr for the LB (Ma\'{i}z-Apell\'{a}niz 2001; Bergh\"{o}fer
and Breitschwerdt 2002; Fuchs et al. 2006).  Of course, it is possible
that the clouds we see at the present time arose from a relatively
recent ejection or tearing away of neutral material from the wall of the
LB  (Breitschwerdt, Freyberg and Egger 2000; Cox and Helenius 2003). 
Also, there is evidence that to some extent the clouds mutually protect
one another from evaporation, as we will consider in more detail later
(\S\ref{hard}).

\section{Pressure Balance Between the LB and Galactic Disk}\label{pbal}

If we now accept the proposition that the internal pressure in the LB is
well below $10^4\,{\rm cm}^{-3}\,$K, and that this thermal pressure
dominates over other forms of pressure (which may not be true), we must
then contemplate a different imbalance, namely, the fact that the total
expected pressure in the Galactic disk that immediately surrounds the LB
should be two to three times larger  (Boulares and Cox 1990).  The LB
has extensions of very low density gas above and below the Galactic
plane that seem limitless; these are called ``chimneys''  (Welsh et al.
1999, 2002) -- see Figs. 5, 7 and 8 of Lallement et al.  (2003).  Thus
the LB is not directly exposed to the usual amount of weight of a
vertical column of gas, but still we must address the issue of what
prevents the LB from being strangled by the higher pressure gas that
surrounds it in the Galactic plane.

Perhaps in this instance, unlike what we found for the CHISM/LB
interface in \S\ref{fail}, we can draw upon magnetic fields to provide
the necessary support for the LB wall.  If a supernova remnant develops
in an initially weakly magnetized medium, the field lines should be
compressed as the remnant expands, creating a magnetic wall at the
perimeter that has a field strength that is strong enough to ward off
any encroachment of the outside material as the remnant cools
radiatively and its thermal pressure drops  (Slavin and Cox 1992;
Balsara, Benjamin and Cox 2001).  Instabilities at the wall that would
start to allow localized intrusions of the outside disk gas would carry
the magnetic field with them and, as a result, create convex field
configurations whose tensions toward the outside could inhibit further
inward motion.

Is there any evidence of an enhanced magnetic field at the boundary of
the LB?  The answer is yes: Andersson \& Potter  (2006) measured the
dispersion of polarization angles of starlight emitted by stars just
beyond the LB wall and combined them with measurements of the velocity
dispersion of the gas and determinations of $n$(H) using C~I
fine-structure level populations.  They then used the
Chandrasekhar-Fermi method to estimate that the field strength is
$8_{-3}^{+5}\,\mu$G, which leads to a pressure $B^2/(8\pi
k)=1.8_{-1}^{+3}\times 10^4\,{\rm cm}^{-3}$K.

\section{Ionization State of the CHISM}\label{ionization}

\subsection{Initial Findings}\label{main_findings}    

One of the early surprises about the ionization of the local gas was
that observations of neutral H and He absorptions to local white dwarfs
indicated that the fractional ionization of helium was about equal to
that of hydrogen  (Kimble et al. 1993a, b; Vennes et al. 1993) or even
greater  (Green, Jelinsky and Bowyer 1990; Dupuis et al. 1995), in spite
of the fact that the recombination coefficient of He with free electrons
is greater than that of H over temperatures of interest.  There are many
sources of UV radiation that can partially photoionize these two
elements in the CHISM.  A dominant source of such photons is the star
$\epsilon$~CMa  (Vallerga and Welsh 1995), by far the brightest star in
the sky in the EUV spectral region.  However this flux, supplemented by
radiation from nearby white dwarfs, only amounts to photoionization
rates $\Gamma({\rm H})=1.56\times 10^{-15}\,{\rm s}^{-1}$ and 
$\Gamma({\rm He})=8.25\times 10^{-16}\,{\rm s}^{-1}$ according to
Vallerga  (1998).  (Vallerga calculated that emissions from unobserved
late-type stars could raise the He ionization rate somewhat, but only by
about 14\%.)  These values are not sufficient to maintain the observed
levels of ionization.

\subsection{Incomplete Recombination?}\label{incomplete}

One might question the validity of the assumption that the ionization of
the CHISM has reached an equilibrium, as did Lyu \& Bruhweiler  (1996). 
They suggested that a shock created by a nearby supernova about 1 to a
few Myr ago might have strongly heated and ionized the gas, leading a
condition where the material is now overionized (for its current
temperature) because the $e$-folding time for He recombination
$[\alpha({\rm He})n(e)]^{-1}=1.8\,{\rm Myr}$, whereas the gas can cool
much more rapidly.  The shock seemed to be the only mechanism that could
work: both Lyu \& Bruhweiler  (1996) and Frisch \& Slavin  (1996)
concluded that the UV flash from any very recent, nearby supernova would
not be able to appreciably ionize the local gas (unless it were very
much closer than the closest known supernova in the constellation of
Vela).

Sofia \& Jenkins  (1998) proposed that a way to test the proposal of
incomplete recombination would be to examine the abundance of Ar~I
relative to H~I in the local gas.  Argon is not likely to be depleted in
the ISM, so measurements of the ratio of neutral forms of these two
elements indicates their relative levels of ionization.  The
recombination rates of Ar and H are about equal, whereas the
photoionization cross section of Ar is much higher than that of H. 
Thus, as Jenkins \& Sofia pointed out, a condition of steady-state
ionization would yield Ar~I/H~I less than the underlying abundance ratio
of these elements (assumed to be solar), and conversely an incomplete
recombination from a much more highly ionized state would yield a ratio
more in accord with the real value of Ar/H.  Later measurements of
$N$(Ar~I) relative to $N$(O~I)\footnote{O~I is a good surrogate for H~I 
because oxygen atoms are not heavily depleted onto dust grains and their
relative ionization is strongly coupled to that of H through a strong
charge-exchange reaction  (Field and Steigman 1971).  Thus, taking
$N$(O~I) and dividing by the solar abundance ratio of O to H should
yield a good estimate for $N$(H~I).} toward nearby white dwarf stars
revealed that Ar~I is indeed deficient  (Lehner et al. 2003), supporting
the notion that the ionization of the local medium is maintained by a
steady-state photoionization.

\subsection{Diffuse Hard Photoionization Flux}\label{hard}

If we now accept the idea that the ionization is in equilibrium, a
possible solution to the apparent excess helium ionization is the
ionizing effect provided by the additional, much harder radiation
emitted by the conductive interface around the CHISM  (Slavin 1989),
supplemented by the soft X-rays from the LB.  Slavin \& Frisch  (2002)
have calculated that these two sources of hard radiation, working in
conjunction with the fluxes from stars, could explain the observed level
of ionization of H and He in the CHISM.

Slavin \& Frisch  (2002) made reasonable allowances for a number of
possible effects that could alter the conduction rate, and hence the
intensity of the emitted radiation, from the local cloud's conductive
interface.  They considered the possible influence of a randomly
oriented (but not tangled) magnetic field that would slightly suppress
the conduction, the effect of conduction saturation  (Cowie and McKee
1977; Dalton and Balbus 1993), and the modification of the cooling rate
due to the partial ionization of H and He.  From the discussion in
\S\ref{lifetimes}, it is clear that the conduction and evaporation rates
depend on the pressure of the external hot gas.  Thus one might suppose
that if this pressure were to be revised downwards, the inferred
ionizing flux from the interface should be lowered as well.  Evidently
this is not the case for the flux calculated by Slavin \& Frisch 
(2002), since their integration of the outward evaporation flow and its
changing state (including the emission rate) started with a boundary
condition that corresponded to the pressure inside the cloud, instead of
imposing the requirement that there be a match to what now seems to be
an unreasonably high pressure for the surrounding hot medium (J.~Slavin,
private communication).  While this may be so, there is still a need to
lower the estimate for the portion of the ionizing flux that comes from
the hot gases within the LB itself.

It is quite possible that the CHISM does not have a well established
conduction and evaporation front because it is shielded from the hot
plasma by other clouds in the local vicinity.  Balbus  (1985) derived
solutions for the conduction flows for complex geometries.  One outcome
of his study was that individual clouds well inside an ensemble of
neighboring clouds should have significantly reduced flows because of
obstruction by neighboring clouds, in a manner analogous to
electrostatics where a conducting body is shielded from external
potential differences by (grounded) nearby bodies that surround it (as
in the extreme case of a Faraday cage).  He expressed a criterion for
the maximum volume filling factor $f$ of the clouds having radii $a$
within a spherical volume of radius $R$,
\begin{equation}\label{fill_fac}
f \ll {2\over 3}~{a^2\over R^2}
\end{equation}
that is required for a cloud in the center to be unaffected by the
shielding.

In order to make use of Eq.~\ref{fill_fac}, we need to know something
about the cloud environment in the local vicinity.  From the appearance
of UV absorption lines for stars out to various distances  (Redfield and
Linsky 2004a) and the detections of astrospheres\footnote{Astrospheres
are structures created by the interaction of a star's stellar wind with
an ambient medium, much like the heliosphere around the Sun. They are
detected by unique L$\alpha$ absorption features that can be seen on top
of the stellar L$\alpha$ emission line, as long as we view the star from
a direction that is generally near the upwind direction for the star's
motion through its surroundings.} around stars in different locations 
(Wood et al. 2005), it seems that the CHISM is near the center of an
ensemble of similar clouds that fills a volume out to a radius of about
10~pc from the Sun.  Beyond that distance, the systems that contain some
neutral hydrogen are much more sparsely distributed.  From the frequency
of appearance of astrospheres, Wood et al.  (2005) estimated that
$f=0.59$.  Redfield \& Linsky  (2008) arrived at a different estimate of
$0.055 < f < 0.19$, based on the coverage of UV absorption systems that
exhibited coherent velocities over patches of the sky.  In either event,
for characteristic cloud sizes of 2~pc or smaller\footnote{In some
instances, what may seem to be multiple clouds may in fact be velocity
bunching within a single, large cloud.  While this may be so, it does
not alter the basic conclusion that significant shielding may still
occur.} within the volume with radius $R=10\,$pc, the inequality shown
in Eq.~\ref{fill_fac} is not satisfied.  Thus, shielding appears to be a
significant effect that can reduce or virtually eliminate the conduction
front around the CHISM, as well as its closest neighboring clouds.

Two different observational outcomes appear to support an absence of
multiple conduction fronts within the ensemble of local clouds, but
still allow for a single front at the edge of the complex.  In a
collisionally ionized medium, five-times ionized oxygen appears mostly
at temperatures of around $3\times 10^5\,$K,\footnote{This value is for
an equilibrium condition.  In evaporation and condensation flows, the
preferred temperature can be somewhat lower or higher because the
temperature changes more rapidly than the time needed to establish an
equilibrium.} a condition that should arise in the outflow of gas from
the surface of an evaporating cloud.  For the first measurement, we note
that for a single direction in the sky, Shelton  (2003) found a
$2\sigma$ upper limit of $800\,{\rm photons~cm}^{-2}{\rm s}^{-1}{\rm
str}^{-1}$ for the brightness of the 1032, 1038\AA\ doublet of O~VI, and
this limit is about consistent with the predicted intensity from a
conduction front of $600-1000\,{\rm photons~cm}^{-2}{\rm s}^{-1}{\rm
str}^{-1}$, depending on the age of the front  (Borkowski, Balbus and
Fristrom 1990).  This limit seems to rule out multiple fronts
(but,again, for only one particular direction).  The second type of
evidence arises from a survey of O~VI absorption features in the spectra
of white dwarf stars within the LB  (Oegerle et al. 2005; Savage and
Lehner 2006)\footnote{The reality of the interstellar O~VI absorptions
has recently been questioned by M.~Barstow and colleagues.  At the time
of writing of this article, the evidence that supports this conclusion
has not yet appeared in the literature, but the article by B.~Welsh in
this volume presents a brief report on the principal claims by this
group.}.  Generally, when detected, the O~VI features exhibited
kinematic properties that were consistent with our looking outwards
through a single evaporation front: the lines had the correct width, and
they had a tendency to show a positive velocity shift relative to
absorptions from C~II that arise from the cloud material.  

Since the cloud ensemble ultimately has an outer boundary that meets the
hot plasma of the LB, the gas within the CHISM will still be exposed to
ionizing radiation from all directions -- the only difference here is
that this diffuse illumination will come from a much greater distance 
($\approx$~10~pc) than the CHISM boundary.  For this reason, the
radiation will be absorbed by gaseous material in the clouds with a
total column density of hydrogen equivalent to about $10^{18}\,{\rm
cm}^{-2}$ [but for a few directions, much more than this value,
according to Wood et al.  (2005)].  A column this large will attenuate
photons at energies just above the He~I ionization limit by about a
factor of 4, but this absorption will diminish for the higher
energies.\footnote{According to Hurwitz, Sasseen \& Sirk  (2005), for a
solar abundance plasma at $T=10^6\,$K, 90\% of the helium-ionizing
photons come from an EUV complex of Fe lines near 180\AA\ (i.e., the
same lines that they were looking for -- see \S\protect\ref{EUV_bkgnd})
-- the calculation of this percentage includes a weighting by the change
of the helium ionization cross section with energy.  At this wavelength,
the attenuation is much less than that near the He ionization edge.  If
Fe is depleted, however, these lines should no longer be as dominant.}

In short, the rate of ionization arising from both the X-ray background
and the conductive interfaces could be lower than the estimate furnished
by Slavin \& Frisch  (2002).  While their investigation was a major step
in the right direction, perhaps some revised thinking on this important
topic is justified by the more recent findings.

\begin{acknowledgements}
The author thanks B.~Welsh for alerting him about the X-ray leak problem
with EUVE.  J. Slavin provided useful advice on conduction fronts.
\end{acknowledgements}

\end{document}